\documentclass[fleqn]{scrartcl}
\usepackage{amsmath, graphicx,indentfirst,amsfonts,amssymb}
\renewcommand{\Im}{\mathop{\mathrm{Im}}\nolimits}
\begin{document}
\begin{center}
\vspace{30pt}
{\Large{\bf Photon-photon interaction and Schwinger process}
}
\vspace{30pt}

K.\,Bulycheva\\

{\it Institute for Theoretical and Experimental Physics, Moscow, Russia

Moscow Institute of Physics and Technology, Dolgoprudny, Russia

bulycheva@itep.ru
}

\vspace{30pt}

S.\,Guts\\

{\it Institute for Theoretical and Experimental Physics, Moscow, Russia

Moscow Institute of Physics and Technology, Dolgoprudny, Russia

guts@itep.ru
}

\vspace{30pt}
\end{center}

\begin{center}
{\bf\abstractname{}}
\end{center}

\begin{abstract}
We consider the tunneling processes induced by two-particle collisions. The
result derived by the thermal bath approach is reproduced by the classical
configurations. The Coulomb and Yukawa corrections for the rate of the process
are found.
\end{abstract}
\pagebreak
\pagestyle{plain}
\section{Introduction}

The problem of the pair creation in the external field has been investigated since historical works of Schwinger \cite{schwinger} and Euler and Heisenberg \cite{euler}. However, this process hasn't been observed experimentally due to huge strengths of fields needed. The pair production rate is exponentially
suppressed when the strength of the electrical field $E$ is lower than $\frac {m_e^2}e$, which is much larger than any available one. The natural way to observe the pair creation in weaker fields is to consider the pair creation induced by external particles, for example by hard photons in the field of high-energy lasers \cite{schutzhold}, \cite{dunne}. This phenomenon is referred to as induced Schwinger process.

There are various types of Schwinger-like processes. They all are closely related to the decay of false vacuum. The decay of the false vacuum in case of scalar fields was studied in \cite{okun}, \cite{callan}, \cite{coleman}, \cite{SelivanovVoloshin},
\cite{Kiselev}. These results can be generalized to the case of electron-positron pair creation \cite{electron}. Monopole decay in constant electrical field appears to be quite similar to induced Schwinger process \cite{gorsky},
\cite{monin_lagr}, \cite{zayakin}, \cite{mo-za}. Schwinger type processes also arise in the weak coupling limit of the string theory \cite{gorsky}, \cite{gorsky2}.

In this paper we consider the Schwinger process induced by several external particles in two-dimensional electrodynamics and two-dimensional scalar theory. We mostly use the techniques developed in \cite{scalar}, \cite{electron}. First we consider the process induced by two external particles. To find the rate of the process we calculate the imaginary part of the four-fold correlator in the momentum representation and see that the result can be generalized to the case of arbitrary number of external particles. Then we use this result to compute the correction to the action caused by the interaction of the wall of the bounce with itself. We see that the consideration of the Coulomb and Yukawa interaction changes the radius of the bounce.

\section{Cross-section of the process enhanced by two particles}
\subsection{Scalar field}
In this paper we focus on two examples of the two-dimensional interacting field theories which allow the processes of the false vacuum decay, namely the theory of two scalar fields, one of which is in quartic potential, and electrodynamics in presence of constant electric field. Both theories are considered in the limit of the thin-wall approximation, so that the wall of the bounce can be represented as a $\delta$-function. The main goal of the analysis is to study the scattering of the particles on the background of the bounce of the false vacuum in the semiclassical limit. We claim that the interaction with external particles significantly changes the rate of the process even when the momenta of the external particles are small.

First we consider the decay of the false vacuum in the scalar theory and reproduce the result of \cite{scalar}. In that work the cross-section of the scattering of the scalar particles at the background of the bubble of the false vacuum was computed. The authors have used the thermal bath approach. We will calculate the same cross-section directly and show that this calculation can be straightforwardly generalized to any number of interacting particles.

We consider the theory of two interacting scalar fields, namely the free massive field $\chi$ and the field $\phi$ in the quartic potential. In the following we study the scattering of the $\chi$-particles.

\begin{equation}
\label{lagrangian}
\mathcal L=\frac 12 \left(\partial_\mu \phi\right)^2+\frac 12 \left(\partial_\mu
\chi\right)^2-\frac 12 m^2 \chi^2 -V(\phi)- V_{int}(\phi, \chi)=\mathcal L_\chi+\mathcal L_\phi+\mathcal L_{int}.
\end{equation}

The quartic potential $V(\phi)$ has two minima which are referred to as the "false" and the "true" vacua,

\begin{equation}
V(\phi)=\frac 14 \lambda \left(\phi^2-\nu^2\right)^2-\frac{\varepsilon}{2\nu}\left(\phi+\nu\right).
\end{equation}

We consider the mass of the $\chi$ field to be much less than the mass of the $\phi$ field and the wall of the bounce to be thin compared to the radius,

\begin{equation}
\label{thin-wall}
m \ll M=\nu\sqrt{\lambda}, \qquad R \gg 1/M.
\end{equation}

In the absence of the $\chi$ field the action of the bounce can be divided into the internal part and the boundary part,

\begin{equation}
\label{action_0}
 S= S_{wall}+S_{int}=\mu L - \varepsilon A, \qquad \mu = \nu^3 \frac{\sqrt{8\lambda}}{3}.
\end{equation}

Here $L$ is the length of the wall of the bounce and $A$ is the area of the bounce. The extremum of the (\ref{action_0}) is reached when the bounce is circular and its radius is given by:

\begin{equation}
R=\frac{\mu}{\varepsilon},
\end{equation}

and the extremized action becomes:

\begin{equation}
S_0=\frac {\pi \mu^2} \varepsilon.
\end{equation}

The rate of the spontaneous decay can be expressed in terms of the path integral \cite{voloshin},

\begin{equation}
\frac \Gamma L=\frac 2{LT} \Im \int \mathcal D \phi e^{-S[\phi]}=\frac \varepsilon {2\pi} \exp\left(-\frac {\pi \mu^2}\varepsilon\right).
\end{equation}

We consider the interaction term in (\ref{lagrangian}) to be linear on $\chi$,

\begin{equation}
\label{interaction}
V_{int}=g \rho(\phi) \chi.
\end{equation}

The exact form of the function $\rho(\phi)$ is not very important, but it must significantly differ from zero only when the field $\phi$ is far from vacuum. The interaction is non-zero at the boundary of the bounce, so the function $\rho$ has the shape of the $\delta$-function of the radial coordinate,

\begin{equation}
\label{rho_def}
\rho^B({\bf r})=\frac g{2\pi R}\delta\left(r-R\right).
\end{equation}

To find the rate of the process induced by two $\chi$ particles we consider the four-fold correlator,

\begin{equation}
\Im \int \mathcal D \chi \mathcal D \phi \frac {\delta^4}{\delta \chi^4} \exp \left( -S[\phi,\chi]\right)=\Im \int \mathcal D \phi \mathcal D \chi \rho^4(\phi)e^{-S[\phi,\chi]},
\end{equation}

which in the leading semiclassical approximation is given by

\begin{multline}
\label{Pi}
\Im \Pi^{(4)}({\bf r_1},{\bf r_2},{\bf r_3})=\Im \int \mathcal Dr \rho^B({\bf r_1})  \rho^B({\bf r_2})  \rho^B({\bf r_3})  \rho^B(0,0) e^{-S[{\bf r}]}=\\
\frac \Gamma {2L} \int  \rho^B({\bf r_1}-{\bf r})  \rho^B({\bf r_2}-{\bf r})  \rho^B({\bf r_3}-{\bf r})  \rho^B({\bf r}) d^2{\bf r}.
\end{multline}

\begin{figure}
\begin{center}
\includegraphics[width=150pt]{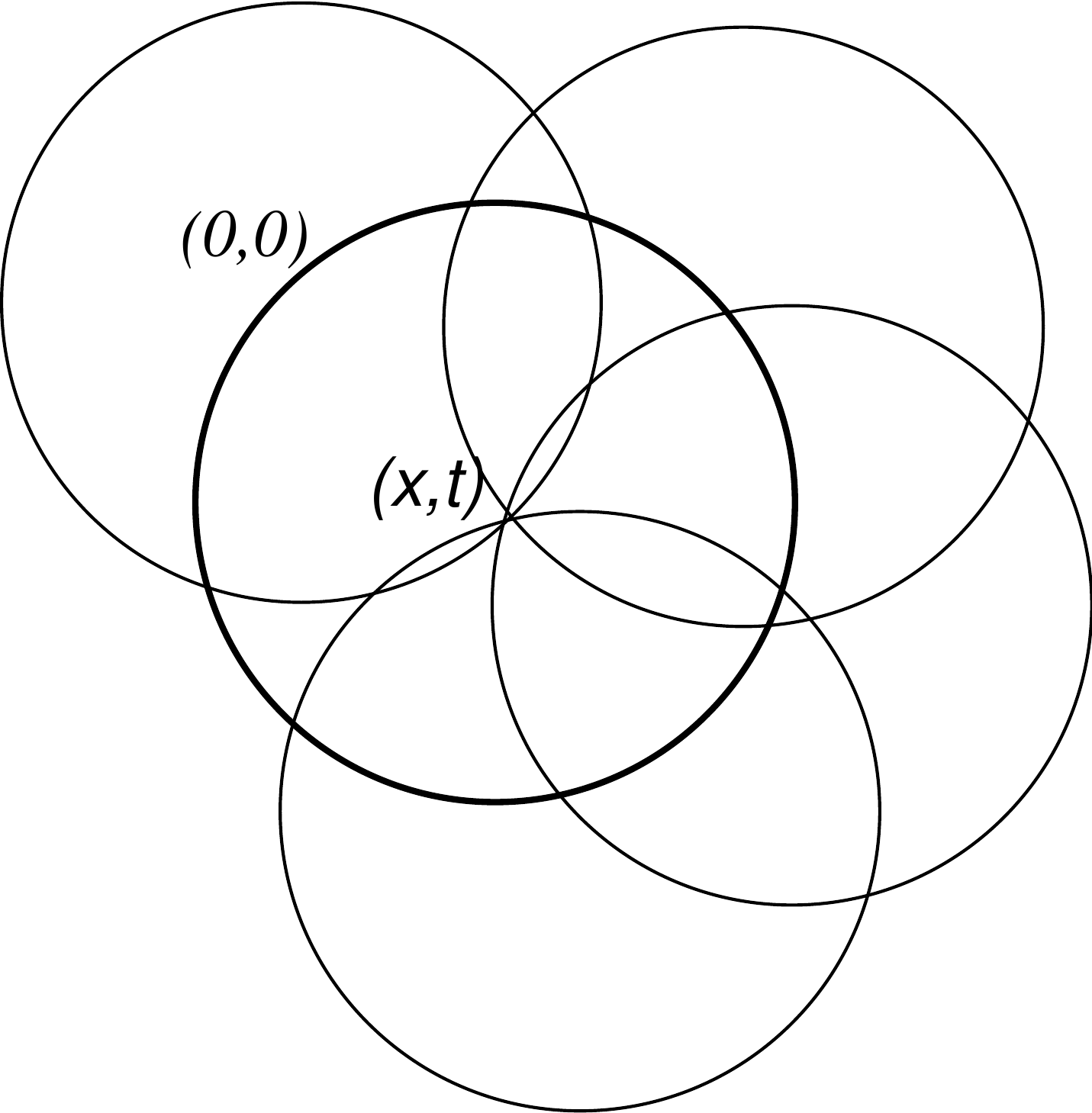}
\end{center}
\caption{The bold circle represents the wall of the bounce, the other four circles represent the densities of $\chi$ particles. The centers of the four circles lie on the bold circle, because the $\delta$-functions force the four circles to intersect in one point.}
\label{corr}
\end{figure}

Geometrically the support of each density $\rho^B$ is a circle with fixed radius. In the first quantized picture the centers of the circles must belong to the boundary of the bounce (see fig. \ref{corr}). This consideration might give an additional condition on relative locations of ${\bf r_i}$; however, this condition is satisfied automatically because the centers of four circles with equal radii intersecting in a single point necessarily belong to the circle of the same radius. This feature of the geometry allows to forget about relative position of the densities and to study the problem in the momentum representation.

First we consider the correlator (\ref{Pi}) in the coordinate form. The integration over ${\bf r}$ gives

\begin{equation}
\label{step1}
\Im \Pi ({\bf r_1}, {\bf r_2}, {\bf r_3})=\frac{\Gamma}{2L}\frac {g^4}{(2\pi R)^4}\frac{4R^2}{r_1\sqrt{4R^2-r_1^2}} \delta\left(r_2-R\right)\delta\left(r_3-R\right).
\end{equation}

Switching to the momentum representation we obtain a simple expression for the four-fold correlator. ${\bf q_i}$ are the momenta of the $\chi$ particles, so that $|{\bf q_i}|=q_i=m$.

\begin{multline}
\label{step2}
\Im \Pi ({\bf q_1},{\bf q_2},{\bf q_3})=\frac{\Gamma}{2L}\int \frac{g^4\delta\left(r_2-R\right)\delta\left(r_3-R\right)}{\pi^2(2\pi R)^2r_1\sqrt{4R^2-r_1^2}} e^{i({\bf q_1 r_1})+i({\bf q_2 r_2})+i({\bf q_3 r_3})}d^2{\bf r_1}d^2{\bf r_2}d^2{\bf r_3}=\\
-\frac {g^4}{(2\pi R)^2}I_0^2(mR)\int \delta\left(r_2-R\right)\delta\left(r_3-R\right)e^{i({\bf q_2 r_2})+i({\bf q_3 r_3})}d^2{\bf r_2}d^2{\bf r_3}=-\frac \Gamma {2L}g^4I_0^4(mR).
\end{multline}

Here the integral representation of the Bessel function was used,

\begin{equation}
\label{bessel_1}
\int\delta\left(r-R\right)e^{i({\bf q r})}d^2{\bf r}=\int e^{iqR\cos\theta}Rd\theta=2\pi R
J_0(qR)=2\pi RI_0(mR),
\end{equation}
\begin{equation}
\label{bessel_2}
\int\frac{e^{i({\bf q r})}}{r\sqrt{4R^2-r^2}}d^2{\bf r}=\int \frac{e^{iqr\cos
\theta}}{\sqrt{4R^2-r^2}}drd\theta=-\pi^2I_0^2(mR).
\end{equation}

We see that the geometry of the problem does not impose additional restrictions on the positions of ${\bf r_i}$. Hence we can first switch to the momentum form of the densities (\ref{rho_def}), and then compute the correlator. This significantly simplifies the calculations.

The density in the momentum representation is as follows,

\begin{equation}
\label{rho_q}
\rho^B({\bf r_1}-{\bf r})\rightarrow\rho^B({\bf q_1},{\bf r})=\int\rho^B({\bf r_1}-{\bf r})e^{i({\bf q_1 r_1})}d^2{\bf r_1}=I_0(mR)e^{i({\bf q_1 r})}.
\end{equation}

The calculation of the correlator in terms of densities (\ref{rho_q}) reduces to the computation of an integral of type (\ref{bessel_1}),

\begin{multline}
\label{momentum}
\Im \Pi ({\bf q_1},{\bf q_2},{\bf q_3})=\frac{\Gamma}{2L} \frac{g^4}{(2\pi R)^4}\int\rho^B({\bf q_1},{\bf r})\rho^B({\bf q_2},{\bf r})\rho^B({\bf q_3},{\bf r})\rho^B({\bf r})d^2{\bf r}=\\\frac \Gamma {2L}\frac {g^3I_0^3(mR)}{2\pi
R}\int\delta\left(r-R\right)e^{i\left(({\bf q_1}+{\bf q_2}+{\bf q_3}),{\bf r}\right)}d^2{\bf r}=-\frac \Gamma {2L}g^4I_0^4(mR),
\end{multline}

and can be easily generalized to a calculation of an $n$-fold correlator,

\begin{equation}
\label{n-fold}
\Im \Pi^{(n)}=-\frac \Gamma {2L} \left(g I_0(mR)\right)^n.
\end{equation}

In the expression (\ref{momentum}) the momentum conservation law was used, $\sum{\bf q_i}=0$ , $|{\bf q_4}|=m$. To reproduce the result of \cite{scalar}, we write the cross-section in terms of the imaginary part of the correlator,

\begin{equation}
\sigma=-\frac{\Im \Pi^{(4)}}{2I}=\frac{\varepsilon}{4\pi}\frac1{2I}(gI_0(mR))^4\exp\left(-\frac{
\pi\mu^2}{\varepsilon}\right),
\end{equation}

where $I=\sqrt{({\bf q_1q_2})^2-m^4}$ is a kinematical invariant. Using (\ref{n-fold}), we can calculate the rate of the process induced by any number of the external particles.

\subsection{Electrodynamics}

Now we move to our second example, namely the electrodynamics in two Euclidean dimensions in presence of a constant electric field. Our goal is to calculate the rate of the pair creation enhanced by two photons. We proceed just the same way as before, i.e. we calculate the imaginary part of the four-fold correlator and argue that this technique allows to calculate the polarization operator with any number of external photons.

The creation of an electron-positron pair in external field can be thought of as the creation of a bubble of the true vacuum in the false vacuum. The action of the configuration consists of a surface term and a boundary term, just as in the scalar case (\ref{action_0}), with the following identification of the parameters:

\begin{equation}
\mu = m,\qquad \varepsilon = eE,\qquad R=\frac{m}{eE},
\end{equation}

where $m$ is the mass of electron. The rate of the spontaneous process is,

\begin{equation}
\label{gamma2}
\frac {\Gamma} {L}=\frac {eE}{2\pi}\exp\left(-\frac{\pi m^2}{eE}\right).
\end{equation}

Applying the technique of calculation of the imaginary part of the four-fold correlator, we find the polarization operator in leading order on $\frac {eE}{m^2}$ (neglecting the spinorial structure of the electron current). Then the electron current is of the form,

\begin{equation}
\label{current}
j^{\mu}({\bf r})=en^{\mu}\delta \left(r-R\right)=e\frac {\epsilon^{\mu\nu}r_\nu}r \delta (r-R),\qquad {\bf n} =\begin{pmatrix}\sin \phi\\-\cos\phi\end{pmatrix},\qquad \mu,\nu=1,2,
\end{equation}

where $\bf n$ is a unit vector tangential to the boundary of the bounce.

We calculate two-fold and four-fold polarization operators as the imaginary parts of the corresponding correlators,

\begin{equation}
\label{ImPi2}
\pi^{\mu \nu}({\bf r_1})=\Im \Pi^{\mu \nu}({\bf r_1})=\frac {\Gamma} {2L} \int
j^{\mu}\left({\bf r_1}-{\bf r}\right)j^{\nu}\left(-{\bf r}\right)d^2{\bf r},
\end{equation}

\begin{multline}
\label{ImPi}
\pi^{\mu \nu\kappa\rho}({\bf r_1}, {\bf r_2}, {\bf r_3})=\Im \Pi^{\mu \nu \kappa \rho}({\bf r_1}, {\bf r_2}, {\bf r_3})=\\
\frac {\Gamma} {2L} \int j^{\mu}\left({\bf r_1}-{\bf r}\right)j^{\nu}\left({\bf r_2}-{\bf r}\right)j^{\kappa}\left({\bf r_3}-{\bf r}\right)j^{\rho}\left(-{\bf r}\right)d^2{\bf r}.
\end{multline}

We switch to the momentum form of the current (\ref{current}),

\begin{multline}
\label{currentOrt}
j^\mu ({\bf r_1}-{\bf r})\rightarrow j^{\mu}({\bf q_1},{\bf r})=e^{i({\bf q_1 r})}\int e \begin{pmatrix} \sin \phi \\ -\cos \phi \end{pmatrix}\delta(r-R) e^{iq_1r\cos(\theta_1-\phi)}rd\phi dr=\\
-2\pi ieRJ_1(q_1R)e^{i({\bf q_1 r})}\begin{pmatrix}\sin\theta_1 \\ -\cos\theta_1\end{pmatrix}=-2\pi ie \frac {\epsilon^{\mu \nu}{q_1^\nu}}{q_1}RJ_1(q_1R)e^{i({\bf q_1 r})}
\end{multline}

Note that here $q=\sqrt {q_x^2+q_t^2}$ is the euclidean absolute value of ${\bf q}$ and is nonzero though the photon is massless. The calculation similar to (\ref{momentum}) leads to the following expressions for the polarization operators:

\begin{equation}
\label{pop_2}
\pi^{\mu\nu}({\bf q})=\frac {\Gamma} {2L} \left(2\pi e R J_1(qR)\right)^2
\left(\delta^{\mu\nu}-\frac{q^\mu q^\nu} {q^2}\right),
\end{equation}

\begin{equation}
\label{pop_4}
\pi^{\mu\nu\kappa\rho}({\bf q_1},{\bf q_2},{\bf q_3},{\bf q_4})=\frac {\Gamma}{2L} \left(\prod_{i=1}^4 \frac
{2\pi e R J_1(q_iR)}{q_i}\right)\left(q_2^\mu q_1^\nu - ({\bf q_1 q_2})\delta^{\mu \nu}\right)\left(q_4^\kappa q_3^\rho - ({\bf q_3 q_4})\delta^{\kappa \rho}\right).
\end{equation}

The latter expression is subject to the condition of the momentum conservation, $\sum {\bf q_i}=0$. As it was expected from (\ref{currentOrt}), it is transverse to respective momenta.
The result (\ref{pop_4}) directly generalizes to any even number of external photons:

\begin{equation}
 \label{pop_n}
\pi^{\mu_1 \mu_2 \ldots \mu_{2n}} ({\bf q_1},{\bf q_2}, \ldots ,{\bf q_{2n}}) = \frac {\Gamma}{2L} \left( \prod_{i=1}^{2n} \frac {2\pi e R J_1(q_iR)}{q_i} \right) \prod_{i=1}^{n} ( q_{2i}^{\mu_{2i-1}} q_{2i-1}^{\mu_{2i}} - ({\bf q_{2i} q_{2i-1}})\delta^{\mu_{2i-1}\mu_{2i}} )
\end{equation}

\begin{figure}
\begin{center}
\includegraphics[width=200pt]{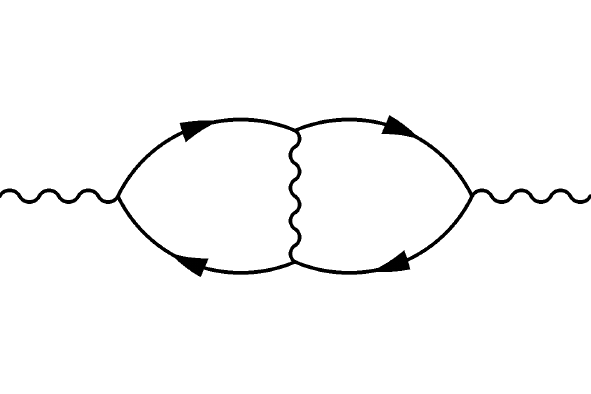}
\end{center}
\caption{The shape of the bounce changes considerably when the interaction with hard photons takes place.}
\label{shape}
\end{figure}

 All the above results were obtained in the limit of soft photons $q\ll m$. Generally speaking, the interaction with external photons considerably changes the form of the bounce, because it acquires cusps at the points where the interaction takes place (see fig. \ref{shape}). This leads to an exponential correction to the rate of the process. However consideration of processes with hard photons results in significant complication of the problem and lies beyond the scope of the current paper.

In the next section we calculate another exponential correction to the action which is caused by interaction of the particles of the bounce with each other. The results (\ref{n-fold}) and (\ref{pop_4}) allow to calculate the rate of the processes in which electrons exchange photons and $\phi$ particles interact by means of $\chi$ field. This corrections are referred to as Coulomb and Yukawa corrections.

\section{Coulomb and Yukawa corrections}

In this section we compute the exponential corrections to the rates of the processes caused by the interaction of the wall of the bounce with itself. A similar quantity was calculated in \cite{affleck} for four-dimensional electrodynamics. Their result for the corrected effective action reads as:

\begin{equation}
S_C^{4d}=\frac {\pi m^2}{eE}-\frac {e^2}4.
\end{equation}

The correction is small in the considered limit of small charges and is independent of the radius of the bounce. Hence in the four-dimensional electrodynamics the Coulomb correction does not change the radius of the bounce. We see below that in the two-dimensional case this is not so.

We calculate analogous corrections using our results (\ref{n-fold}) and (\ref{pop_4}) for correlators. Using the fact that the correlators in the momentum representation are products of uniform multiples we can calculate contribution of any number of particles interacting with bounce.

\begin{figure}
\begin{tabular}{ccccccccc}
\begin{tabular}{c} \includegraphics[scale=.3]{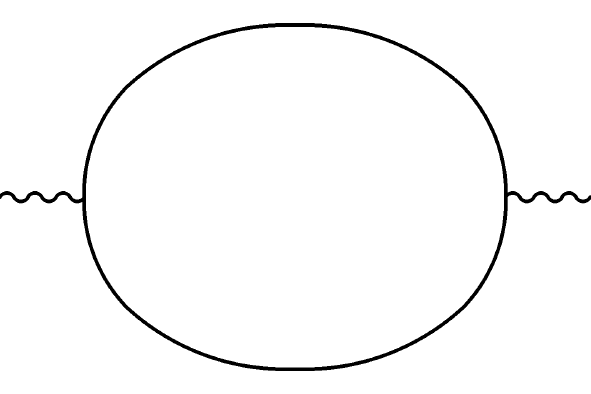} \end{tabular} & + & \begin{tabular}{c} \includegraphics[scale=.3]{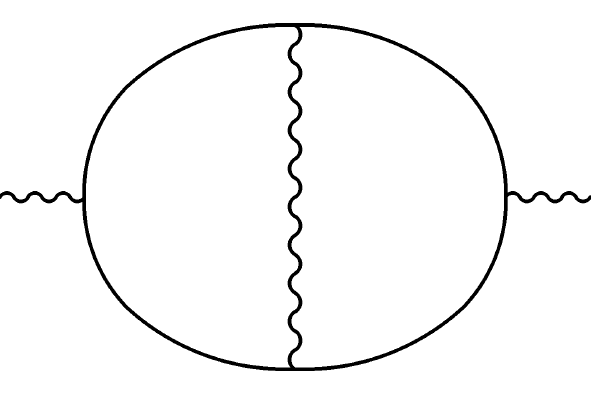} \end{tabular} & + & \begin{tabular}{c} \includegraphics[scale=.3]{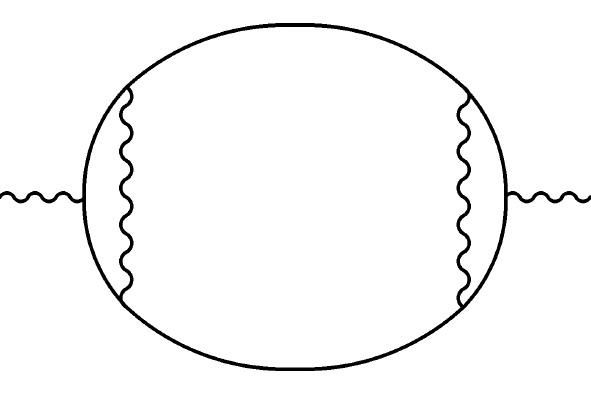} \end{tabular} & + & \begin{tabular}{c}  \includegraphics[scale=.3]{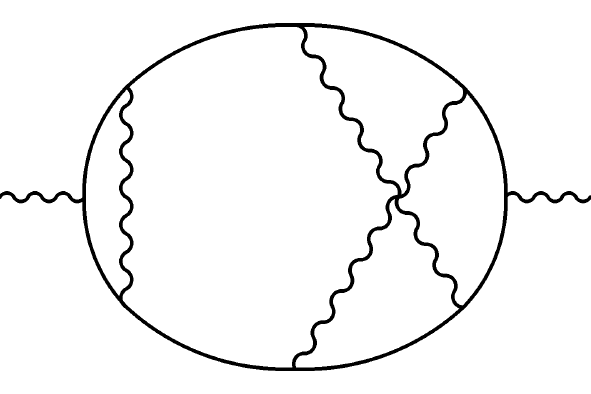} \end{tabular}&+&\ldots\\ $\pi^{\mu \nu}_{0}$ & \qquad & $\pi^{\mu \nu}_{1}$ & \qquad & $\pi^{\mu \nu}_{2}$ & \qquad & $\pi^{\mu \nu}_{3}$&\qquad

\end{tabular}
\caption{To obtain the Coulomb correction for polarization operator, the interaction with arbitrarily many photons is to be studied.}\label{coulomb}
\end{figure}

We begin with the Coulomb correction for the rate of processes in two-dimensional electrodynamics. To calculate the $n$-photon contribution, we consider the $(2n+l)$-fold correlator, where $l$ is the number of external photons. Then we integrate over the momenta of $n$ internal photons with their propagators. To find the full Coulomb correction we have to sum up the contributions of any number of photons taking into account combinatorial factors (see fig. \ref{coulomb}).

First we find one-photon contribution. For simplicity we consider the width of the spontaneous decay, though the case of induced decay is absolutely analogous. We integrate two-photon polarization operator (\ref{pop_2}) with the propagator of the photon.

\begin{multline}
\label{coul}
\frac{\Gamma_1}{2L}= \frac 12 \int_0^\infty \pi^{\mu\nu}({\bf q}) \frac{\delta^{\mu\nu}}{q^2}{\frac {d^2q}{(2\pi)^2}} = \\
= \frac 12 \frac{ \Gamma_{}} {2L} \int_0^\infty \left(2 \pi e R J_1(qR)\right)^2
\left(\delta^{\mu\nu}-\frac{q^\mu q^\nu} {q^2}\right)\frac
{\delta^{\mu\nu}}{q^2}{\frac {d^2q}{(2\pi)^2}}=\frac {\Gamma} {2L} \frac {\pi R^2e^2}2.
\end{multline}

Here $\Gamma_n$ is the $n$-photon correction to the width. The calculation (\ref{coul}) contains a certain subtlety. As we have argued above our results are valid only in case of small photon momenta $q$, but the integration in (\ref{coul}) is over all values of $q$. However the integrand significantly differs from zero only in area $qR \sim 1$, so the contribution from hard photons is negligible.

Corrections $\Gamma_n$ for $n\ge 2$ are calculated  in a similar fashion with $n$-photon polarization operator (\ref{pop_n}) and $n$ photon propagators. As we see on example of $\Gamma_2$,
integrals over momenta of different photons are independent from each other:

\begin{multline}
 \frac{\Gamma_2}{2L} = \frac 14 \int_0^\infty \int_0^\infty \pi^{\mu\nu\kappa\rho}({\bf q},{\bf -q},{\bf p},{\bf -p}) \frac {\delta^{\mu\nu}}{q^2} \frac {\delta^{\kappa\rho}}{p^2}
 \frac {d^2q}{(2\pi)^2} \frac {d^2p}{(2\pi)^2} = \\
= \frac 14  \frac{ \Gamma_{}} {2L} \left( \int_0^\infty \left(2 \pi e R J_1(qR)\right)^2
\left(\delta^{\mu\nu}-\frac{q^\mu q^\nu} {q^2}\right)\frac
{\delta^{\mu\nu}}{q^2}{\frac {d^2q}{(2\pi)^2}} \right)^2 = \frac {\Gamma} {2L} \left(\frac {\pi R^2e^2}2\right)^2.
\end{multline}

The same thing happends for arbitrary $n$, resulting in $n$-th order correction

\begin{equation}
 \Gamma_n = \Gamma \left(\frac{\pi R^2 e^2}2\right)^n.
\end{equation}

Summing up the contributions of any number of photons and taking into account the combinatorial factor $\frac 1 {n!}$ we find the corrected width:

\begin{equation}
\Gamma_{C}=\sum_{n=0}^{\infty} \frac 1{n!} \Gamma_n= \Gamma \exp \left(\frac{\pi R^2 e^2}2\right).
\end{equation}

All the polarization operators acquire the same exponential factor; for example, the corrected polarization operator corresponding to the process induced by a single photon reads as:

\begin{equation}
\pi^{\mu\nu}_{C}(q)=\pi^{\mu\nu}(q)\exp \left(\frac{\pi R^2 e^2}2\right).
\end{equation}

The exponential factor can be interpreted as an additional term in the effective action:

\begin{equation}
S_C=2\pi mR - \pi eE R^2-\frac 12 \pi R^2 e^2.
\end{equation}

The extremization of the action gives the radius of the process with the Coulomb interaction present:

\begin{equation}
R_C=\frac {2m R}{e(2E+e)}
\end{equation}

We see that the Coulomb interaction decreases the radius of the bounce; however the change is small in the limit $e \ll E$.

In the same way we find the Yukawa correction for the false vacuum decay in scalar theory. The two-fold correlator integrated with the propagator of the $\chi$ particle reads as:

\begin{equation}
\label{yuk}
\Gamma_{1}=\frac 12 \frac \Gamma {2L}\int_0^\infty g^2 J_0(qR)^2
\frac{qdq}{q^2+m^2}=\frac \Gamma {2L} \frac {g^2}2 I_0(mR)K_0(mR).
\end{equation}

The width acquires exponential correction which again depends on the radius:

\begin{equation}
\label{gamma_yuk}
\Gamma_{Y}= \frac{\varepsilon}{4\pi} \exp\left(-2\pi\mu R + \pi \epsilon R^2+ \frac{g^2}2 I_0(mR)K_0(mR)\right).
\end{equation}

The exponential correction $I_0\left(mR\right)K_0\left(mR\right)$ behaves like a logarithm for small argument. However, this is not an unusual property for a two-dimensional theory to have infrared logarithmic divergences. Bessel functions often appear in answers for two dimensions (see e.g. \cite{Gorsky_diver}). The correction is significant in the area of small $m$ where the distortion of the shape of the bounce is negligible. When $m > \mu$ the change in the shape is the main source of the correction.

The correlators for the induced processes acquire the same exponential factor. This factor can be interpreted as an additional term in the action too,

\begin{equation}
S_Y=-2\pi\mu R + \pi \epsilon R^2+ \frac{g^2}2 I_0(mR)K_0(mR),
\end{equation}

and this means that the Yukawa interaction changes the radius of the bounce too. The exact value of the radius can be found as the solution of the extremization condition.

\section{Conclusions}

In this work the induced Schwinger process was studied by purely semicalssical means. We calculated the cross-section for two-photon interaction at the background of the bounce as leading term in $\frac {eE}{m^2}$ expansion which is applicable for the electric field strength much smaller than $E_{crit}$. The process was studied for soft photons ($\omega \ll m_e$). In case of hard photons the cross-section acquires exponential correction because of the change in the shape of the bounce. We do not investigate this correction because the increase of the momenta of the photons leads to considerable complication of the setting.

The calculation of the rate of the two-photon interaction easily generalizes to the case of the process induced by arbitrarily many photons. Extremely simple structure of the corresponding correlator allowed us to compute the correction to the action which is caused by interaction of the wall of the bounce with itself. This correction depends on the radius and therefore results in the change of the radius of the bounce. So we have seen that the Coulomb interaction decreases the radius of the bounce. Similar calculation was carried out for the scalar fields. In both cases the corrections are significant in the limit of soft particles where the distortion of the bounce is negligible.

The authors are especially grateful to A.\,Gorsky for initiating the work and for useful and productive discussions. The work was supported by grants RFBR 12-02-00284-a and RFBR-CNRS 12-02-91052 as well as Dynasty Foundation stipend program.

\end{document}